%% file: MuonStripsNIM.tex
\documentclass[5p]{elsarticle}

\usepackage[utf8]{inputenc}  
\usepackage[T1]{fontenc}       
\usepackage{lineno}
\modulolinenumbers[5]
\usepackage{xspace}            
\usepackage{slashed}
\usepackage{amsmath}           
\usepackage[labelformat=simple]{subcaption}
\usepackage{txfonts}
\usepackage{soul}

\RequirePackage{multirow}
\RequirePackage[colorlinks,breaklinks,citecolor=blue,urlcolor=blue,linkcolor=blue]{hyperref}
\input{definitions}

\input{heppennames2}
\journal{Nuclear Instruments and Methods in Physics Research Section A}

\begin{document}

\begin{frontmatter}
\title{Time and Position Resolution of the Scintillator Strips for a Muon System at Future Colliders}

\author[fnal]{Dmitri Denisov}
\author[ihep]{Valery Evdokimov}
\author[vinca]{Strahinja Lukić\corref{corrauth}}

\address[fnal]{Fermilab, Batavia IL, USA}
\address[ihep]{Institute for High Energy Physics, Protvino, Russia}
\address[vinca]{Vinča Institute, University of Belgrade, Serbia}
\cortext[corrauth]{Corresponding author}
\ead{slukic@vinca.rs}

\begin{abstract}
Prototype scintilator+WLS strips with SiPM readout for a muon system at future colliders were tested for light yield, time resolution and position resolution. Depending on the configuration, light yield of up to 36 photoelectrons per muon per SiPM has been observed, as well as time resolution of \besttime and position resolution along the strip of \bestpos.
\end{abstract}


\end{frontmatter}


\input{IntroNIM.tex}
\input{Setup.tex}

\section{Photon sensitivity and cross-talk calibration}
\label{sec:calibration}
\input{LightCalib.tex}

\section{Tested configurations}
\label{sec:configurations}
\input{Configurations.tex}

\section{Data Analysis}
\label{sec:analysis}
\input{AnalysisNIM.tex}

\input{Results.tex}

\input{Conclusions.tex}

\input{Acknowledgments.tex}

\bibliography{}

\end{document}

%% file: definitions.tex
\newif\ifmarkup

\newcommand{\pT}{\ensuremath{p_\text{T}}\xspace}
\newcommand{\epem}{\ensuremath{\Pep\Pem}\xspace}   
\newcommand{\mumu}{\ensuremath{\PGm\PGm}\xspace}   

\newcommand{\qqbar}{\ensuremath{\PQq\PAQq}\xspace} 
 
\newcommand{\sone}{\textsf{S1}\xspace}
\newcommand{\stwo}{\textsf{S2}\xspace}
\newcommand{\sthr}{\textsf{S3}\xspace}
\newcommand{\sfour}{\textsf{S4}\xspace}

\newcommand{\sipm}{\textsf{SiPM}\xspace}
\newcommand{\sipmo}{\textsf{SiPM1}\xspace}
\newcommand{\sipmt}{\textsf{SiPM2}\xspace}

\newcommand{\cfa}{configuration $A$\xspace}
\newcommand{\cfb}{configuration $B$\xspace}
\newcommand{\cfc}{configuration $C$\xspace}
\newcommand{\cfd}{configuration $D$\xspace}

\newcommand{\bestpos}{\ensuremath{7.7\unit{cm}}\xspace}
\newcommand{\besttime}{\ensuremath{0.45\unit{ns}}\xspace}
\newcommand{\bicron}{Bicron\textsuperscript{\textcopyright} 404A\xspace}

\newcommand{\bicronwls}{Bicron\textsuperscript{\textcopyright} BCF-92\xspace}
\newcommand{\tyvek}{Tyvek\textsuperscript{\textregistered}\xspace}
\newcommand{\tedlar}{Tedlar\textsuperscript{\textregistered}\xspace}
\newcommand{\hamamatsu}{Hamamatsu S10931-050P\xspace}
\newcommand{\dzero}{\ensuremath{\text{D}\slashed{0}}\xspace}

\newcommand{\unit}[1]{\ensuremath{\;\mathrm{#1}}}

\newcommand{\singleplotwid}{0.47\textwidth}

%% file: heppennames2.tex
\def\fileversion{1.5}
\def\filedate{2011/06/05}
\RequirePackage{hepparticles}
\RequirePackage{xspace}
\RequirePackage{amsmath}

    \DeclareFontFamily{U}{fsy}{}
      \DeclareFontShape{U}{fsy}{m}{n}{<->s*[.9]psyr}{}
    \DeclareSymbolFont{ugrf@m}{U}{fsy}{m}{n}
  \DeclareMathSymbol{\upalpha}{\mathord}{ugrf@m}{`a}
  \DeclareMathSymbol{\upbeta}{\mathord}{ugrf@m}{`b}
  \DeclareMathSymbol{\upgamma}{\mathord}{ugrf@m}{`g}
  \DeclareMathSymbol{\updelta}{\mathord}{ugrf@m}{`d}
  \DeclareMathSymbol{\upepsilon}{\mathord}{ugrf@m}{`e}
  \DeclareMathSymbol{\upzeta}{\mathord}{ugrf@m}{`z}
  \DeclareMathSymbol{\upeta}{\mathord}{ugrf@m}{`h}
  \DeclareMathSymbol{\uptheta}{\mathord}{ugrf@m}{`q}
  \DeclareMathSymbol{\upiota}{\mathord}{ugrf@m}{`i}
  \DeclareMathSymbol{\upkappa}{\mathord}{ugrf@m}{`k}
  \DeclareMathSymbol{\uplambda}{\mathord}{ugrf@m}{`l}
  \DeclareMathSymbol{\upmu}{\mathord}{ugrf@m}{`m}
  \DeclareMathSymbol{\upnu}{\mathord}{ugrf@m}{`n}
  \DeclareMathSymbol{\upxi}{\mathord}{ugrf@m}{`x}
  \DeclareMathSymbol{\uppi}{\mathord}{ugrf@m}{`p}
  \DeclareMathSymbol{\uprho}{\mathord}{ugrf@m}{`r}
  \DeclareMathSymbol{\upsigma}{\mathord}{ugrf@m}{`s}
  \DeclareMathSymbol{\uptau}{\mathord}{ugrf@m}{`t}
  \DeclareMathSymbol{\upupsilon}{\mathord}{ugrf@m}{`u}
  \DeclareMathSymbol{\upphi}{\mathord}{ugrf@m}{`f}
  \DeclareMathSymbol{\upchi}{\mathord}{ugrf@m}{`c}
  \DeclareMathSymbol{\uppsi}{\mathord}{ugrf@m}{`y}
  \DeclareMathSymbol{\upomega}{\mathord}{ugrf@m}{`w}
  
  \DeclareMathSymbol{\upvartheta}{\mathord}{ugrf@m}{`J}
  \DeclareMathSymbol{\upvarpi}{\mathord}{ugrf@m}{`v}

  \DeclareMathSymbol{\upvarphi}{\mathord}{ugrf@m}{`j}
  \DeclareMathSymbol{\Upgamma}{\mathord}{ugrf@m}{`G}
  \DeclareMathSymbol{\Updelta}{\mathord}{ugrf@m}{`D}
  \DeclareMathSymbol{\Uptheta}{\mathord}{ugrf@m}{`Q}
  \DeclareMathSymbol{\Uplambda}{\mathord}{ugrf@m}{`L}
  \DeclareMathSymbol{\Upxi}{\mathord}{ugrf@m}{`X}
  \DeclareMathSymbol{\Uppi}{\mathord}{ugrf@m}{`P}
  \DeclareMathSymbol{\Upsigma}{\mathord}{ugrf@m}{`S}
  \DeclareMathSymbol{\Upupsilon}{\mathord}{ugrf@m}{`U}
  \DeclareMathSymbol{\Upphi}{\mathord}{ugrf@m}{`F}
  \DeclareMathSymbol{\Uppsi}{\mathord}{ugrf@m}{`Y}
  \DeclareMathSymbol{\Upomega}{\mathord}{ugrf@m}{`W}

\let\Xspace\xspace

\DeclareRobustCommand{\PZ}{\HepParticle{Z}{}{}\Xspace} 

\DeclareRobustCommand{\Pe}{\HepParticle{e}{}{}\Xspace} 
\DeclareRobustCommand{\Pem}{\HepParticle{\Pe}{}{-}\Xspace} 
\DeclareRobustCommand{\Pep}{\HepParticle{\Pe}{}{+}\Xspace} 

\DeclareRobustCommand{\PGm}{\HepParticle{\upmu}{}{}\Xspace} 

\DeclareRobustCommand{\PQq}{\HepParticle{q}{}{}\Xspace} 

\DeclareRobustCommand{\PAQq}{\HepAntiParticle{\PQq}{}{}\Xspace} 

%% file: IntroNIM.tex
\section{Introduction}
\label{sec:intro}

Several concepts of future colliders, including \Pep\Pem colliders, are currently under study for the next generation of particle physics experiments \cite{ILCTDR13, CLICCDR, FCClongdoc, CEPC-SppC}. Due to the well-defined initial state of the interactions, low backgrounds and radiation levels, \epem colliders are an attractive option for precision measurements to test various theoretical extensions of the Standard Model in the areas where the predictions of the beyond Standard Model theories differ by a few percent, such as in the Higgs sector.

The detector concepts for the future \epem colliders have been developed to a high level of detail over the past decade. Since the publication of the Letters of Intent of the two major concepts, the Silicon Detector (SiD) \cite{SiDLoI} and the International Large Detector (ILD) \cite{ILDLoI}, numerous technical details have been specified to an advanced level. R\&D prototypes of inividual subsystems reach levels of complexity involving hundreds of thousands of readout channels (See e.g.\ Refs.\ 
\cite{DHCAL-proto, SDHCAL-proto}).

However, for the muon systems relatively few specific details are developed, and few experimental tests of detection technologies have been performed. The muon system is envisioned as several layers of position-sensitive detectors embedded in the iron flux-return yoke of the solenoidal magnet.  The role of the muon system at an \epem collider is primarily the identification of muons and track matching to the central tracker, besides serving as the tail catcher for the hadronic showers that penetrate beyond the hadron calorimetry. Examples of previous experimental studies dedicated to the muon-system include tests of a similar detection technique as presented here, but focusing on light yield and attenuation \cite{Balagura06}, and beam tests of a multi-layer prototype devoted to a study of the improvement of energy resolution of a hadronic calorimeter by using the muon system as tail catcher \cite{Adloff12}. The analogue hadronic calorimeter developed and prototyped by the CALICE collaboration uses square scintillator tiles in sizes ranging from $3\times3\unit{cm^2}$ to $12\times12\unit{cm^2}$ with WLS fibers and SiPM readout with the aim of reconstructing the hadronic showers with optimal energy resolution \cite{AHCAL10}. A detailed study of the detection technique similar to the one presented here, focusing on light yield and attenuation for excellent MIP detection efficiency in very long strips for long-baseline neutrino detectors is presented in Ref.\ \cite{Mineev11}. 

\begin{figure}
\centering
   \includegraphics[width=\singleplotwid]{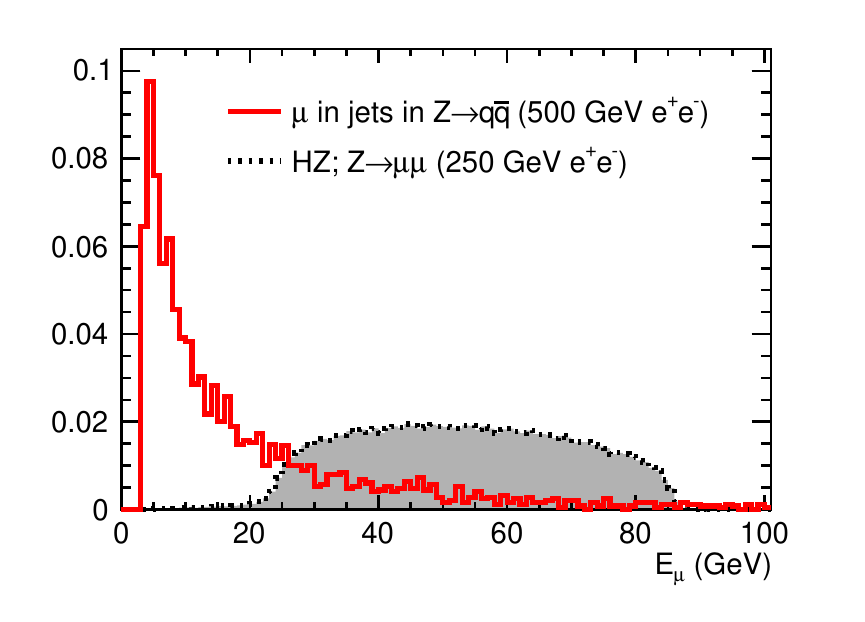}
   \caption{\label{fig:muSpectra} (Color online) Muon energy distribution for muons with sufficient $\pT$ to reach the muon system. Solid line: muons in jets in the $\epem\to\PZ\to\qqbar$ process at a 500 GeV \epem collider. Dashed line: muons from the $\PZ\to\mumu$ decay in the Higgsstrahlung process at a 250 GeV \epem collider.}
\end{figure}

The achievable precision of track matching is limited by the multiple scattering in the detector components before the muon system. The effect of the multiple scattering on track matching can be estimated using the formula by Highland \cite{Highland}. The total thickness of material in the radial direction between the central tracker and the muon system corresponds to about 150-300 radiation lengths, depending on the polar angle. Muons in jets, if they have sufficient $\pT$ to reach the muon system, typically have energies below 10~GeV. Muon spectrum in the process $\epem\to\PZ\to\qqbar$ is shown in Fig.\ \ref{fig:muSpectra} as an example. At such energies, the contribution of the multiple scattering to the smearing of the muon system track position at the first muon-system layer is 5~cm or more. 
Tracks of higher-energy muons, such as those coming from $\PZ\to\mumu$ decay, are less disturbed by the multiple scattering. Such relatively isolated muons are, however, less challenging for track matching in comparison to muons in jets. Fig.\ \ref{fig:muSpectra} shows the example of energy distribution of muons from the $\PZ$ boson decay in the Higsstrahlung process at a 250~GeV \epem collider. 

The total area of the muon detectors to be instrumented with sensitive layers is several thousand square meters. Besides, the iron yoke presents an environment difficult to access for maintenance. For these reasons, economic solutions for a robust and reliable large-area detector are important.

Occupancies in the muon system are moderate at \epem colliders, except in the endcap region of a CLIC collider \cite{Kraaij11}. This allows to consider a strip geometry for the sensitive layers in order to limit the number of readout channels. A promising option consists of scintillator strips with WLS fibers and SiPM readout \cite{ILCTDR4}. In such a system, the coordinates of the muon track are reconstructed using the observables such as the position of the strip hit by a passing muon and the signal time difference $\Delta t$ between the ends of the strip to measure position along the strip. If a muon system has strip orientations alternating by 90 degrees in neighboring layers, the muon track can be reconstructed using only the postions of the strips that fired. In this case, the measurement of the position along each strip using $\Delta t$ may serve to improve the precision of the track fit and to resolve ``ghosts'' arising at the intersections of strips hit by different particles. In the case when perpendicular orientation of strips is not feasible for access to the ends of the strips, time difference remains the only source of information on longitudinal position.

This article is the first in a series devoted to the study of the time resolution and the position resolution achievable from the time difference between the ends of scintillator strips with WLS fibers and SiPM readout. The measurements described in this paper have been performed using cosmic muons at the location of the \dzero assembly building at the Fermi National Accelerator Laboratory, Batavia, USA, at the elevation of 220~m above sea level. The local cosmic muon fluence has been measured previously by the MicroBooNE collaboration to be $\sim 100 \unit{m}^{-2} \unit{s}^{-1}$, with a peak energy between 1 and 2~GeV \cite{Woodruf14}. 

\hamamatsu SiPMs with a sensitive area of $3\times3\unit{mm}^2$ and 3600 pixels each were used for the tests \cite{hamamatsu}. Various scintillators and fibers were used as described below.

Section \ref{sec:setup} describes the measurement setup, Sec.\ \ref{sec:calibration} describes the amplitude and cross-talk calibration, Sec.\ \ref{sec:configurations} describes the tested scintillator strip -- WLS fiber configurations, Sec.\ \ref{sec:analysis} gives details of the data analysis, in Sec.\ \ref{sec:results} measurement results are given and the conclusions are given in Sec.\ \ref{sec:conclusions}.

%% file: Setup.tex
\section{Measurement Setup}
\label{sec:setup}

\begin{figure*}
\centering
\includegraphics[width=.85\textwidth]{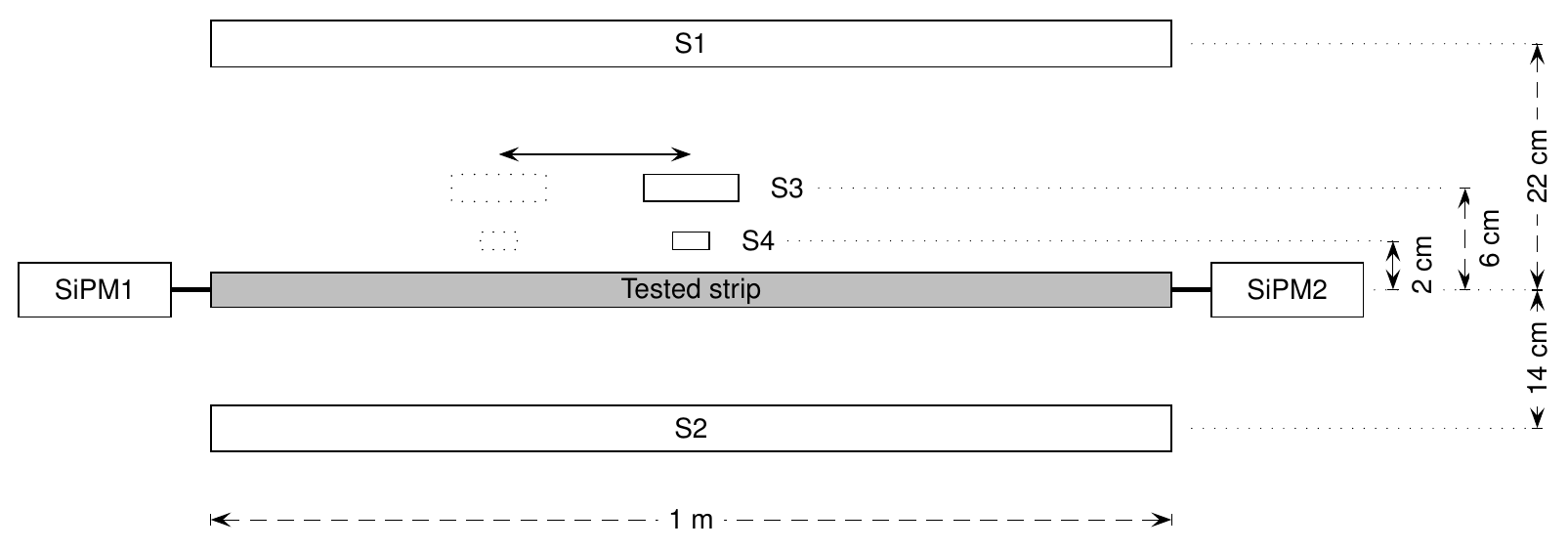}
\caption{\label{fig:setuppads}Schematic of the test setup. \sone and \stwo are scintillation counters with vacuum PMTs positioned above and below the tested strip. \sthr and \sfour are small-area scintillators with vacuum PMT used to select events where the muon hits specific location along the tested strip. The location of \sthr and \sfour w.r.t the tested strip was changed from run to run, keeping the relative position of \sthr and \sfour always the same. \sipmo and \sipmt represent the SiPMs connected to the respective ends of the WLS fibers of the tested strip.}
\end{figure*}

The setup that was used for the measurements is shown in Fig.\ \ref{fig:setuppads}. It was designed to detect cosmic muons by coincidence between vertically arranged scintillation counters. \sone and \stwo are plastic scintillation counters located above and below the tested strip, each of them 1~m long, 10~cm wide and 1~cm thick. \sthr is a 1~cm thick scintillation counter with an area of $10\times15\unit{cm}$, oriented with its 10~cm side along the tested strip located at 6~cm vertical distance from the tested strip. \sfour is a 40~cm long, 2.7~cm wide and 1.2~cm thick scintillation counter oriented across the tested strip and located at 2~cm vertical distance from the tested strip. The counters \textsf{S1-4} were read out using vacuum photomultiplier tubes (PMT). \sipmo and \sipmt denote the SiPMs connected to the respective ends of the WLS fiber of the tested strip. The length of the tested strips is 1~m.

Coincidence between \sone, \stwo and \sthr was used as the trigger, signalling the passage of a muon. The signal from \sthr was delayed by 20~ns with respect to the signals from \sone and \stwo, so that the trigger signal is always formed at the rising edge of the \sthr signal.

The counters \mbox{\sthr} and \mbox{\sfour} were used to restrict the location of the muon to an area smaller than the expected position resolution of the tested strip. The location of \mbox{\sthr} and \mbox{\sfour} along the tested strip was changed from run to run, keeping the relative position of \mbox{\sthr} and \mbox{\sfour} always the same. The distance from the axis of the tested strip to the PMTs of \mbox{\sthr} and \mbox{\sfour} was kept constant to ensure a stable time reference for the measurement of the time resolution of the tested strip. The reason for using two counters for the location restriction was the yet unknown resolution of the tested strips. The counter \mbox{\sfour} provided precise location at the cost of slow counting due to the small intersection area with the tested strip. The counter \mbox{\sthr} provided better counting rate, but introduced an uncertainty of $\pm 5\unit{cm}$ on the longitudinal position. In the offline analysis it was established that the number of hits in the intersection with the counter \mbox{\sfour} with 7-12 hours per point was sufficient, and that the location precision provided by \mbox{\sfour} was necessary to accurately measure the position resolution of the best tested strip. Therefore in the final analysis, the presence of the signal in \mbox{\sfour} was required for event selection.

When measuring properties close to either end of the tested strip, the counters \sone and \stwo were moved along the axis in order to cover locations up to at least 20~cm beyond the end of the tested strip. This was done in order to prevent loss of the muon flux, which would have caused a loss of statistic and a muon position bias at these points.

A CAMAC system with a LeCroy 2249A 12-input charge-sensitive ADC \cite{LeCroyADC} and a LeCroy 2228A 8-input TDC \cite{LeCroyTDC} was used to digitize the amplitude and the arrival time of the signals. The data collection was performed and monitored from a PC with USB connection to the CAMAC Crate controller of type CC-USB by Wiener Plein\&Baus, using custom-made software \cite{WienerDAQ}.

The signals from \sfour, \sipmo and \sipmt were recorded. Each of the signals to be recorded was first split into the time- and the amplitude detection channels using linear fan-in fan-out modules. The time signals were processed using constant threshold discriminators. The time signals were delayed by $\sim 50\unit{ns}$ and digitized by the TDC CAMAC module using the trigger signal as the start. The amplitude signals were delayed by $\sim 40 \unit{ns}$ and digitized by the ADC CAMAC module, using gate generated by the trigger signal. 

A modified setup was used for the measurement of the light yield per muon (Sec.\ \ref{sec:configurations}) and of the attenuation length (Sec.\ \ref{sec:attenuation}). In the modified setup the counters \sthr and \sfour were replaced by a 60~cm long, 2.7~cm wide and 1.2~cm thick \textit{reference} strip counter parallel to the tested strip, located at a 2.5~cm vertical distance, and read out at both ends with vacuum PMTs for muon track longitudinal position measurement. The time resolution of the strip counter is similar to the \sfour counter. The center of the reference strip was positioned to match the center of the tested strip. Using the reference strip we collected required statistics in a reasonable time covering a range of positions along the tested strip. A detailed description of both setups and of all performed measurements is given in Ref.\ \cite{arxivnote}.

The tested strips described here are shorter than what is planned for the muon system at future colliders, while facilitating the prototype studies. In longer strips the signal undergoes more attenuation. The measurement of the attenuation length is described in Sec.\ \ref{sec:attenuation}.

%% file: LightCalib.tex
ADC scale calibration in number of photoelectrons detected was performed by illuminating SiPM with short LED pulses. The driving voltage for the LED had a triangular pulse shape. The stability of the amplitude was monitored by recording an inverted driving signal in a separate ADC channel.

\begin{figure}
\centering
\includegraphics[width=\singleplotwid]{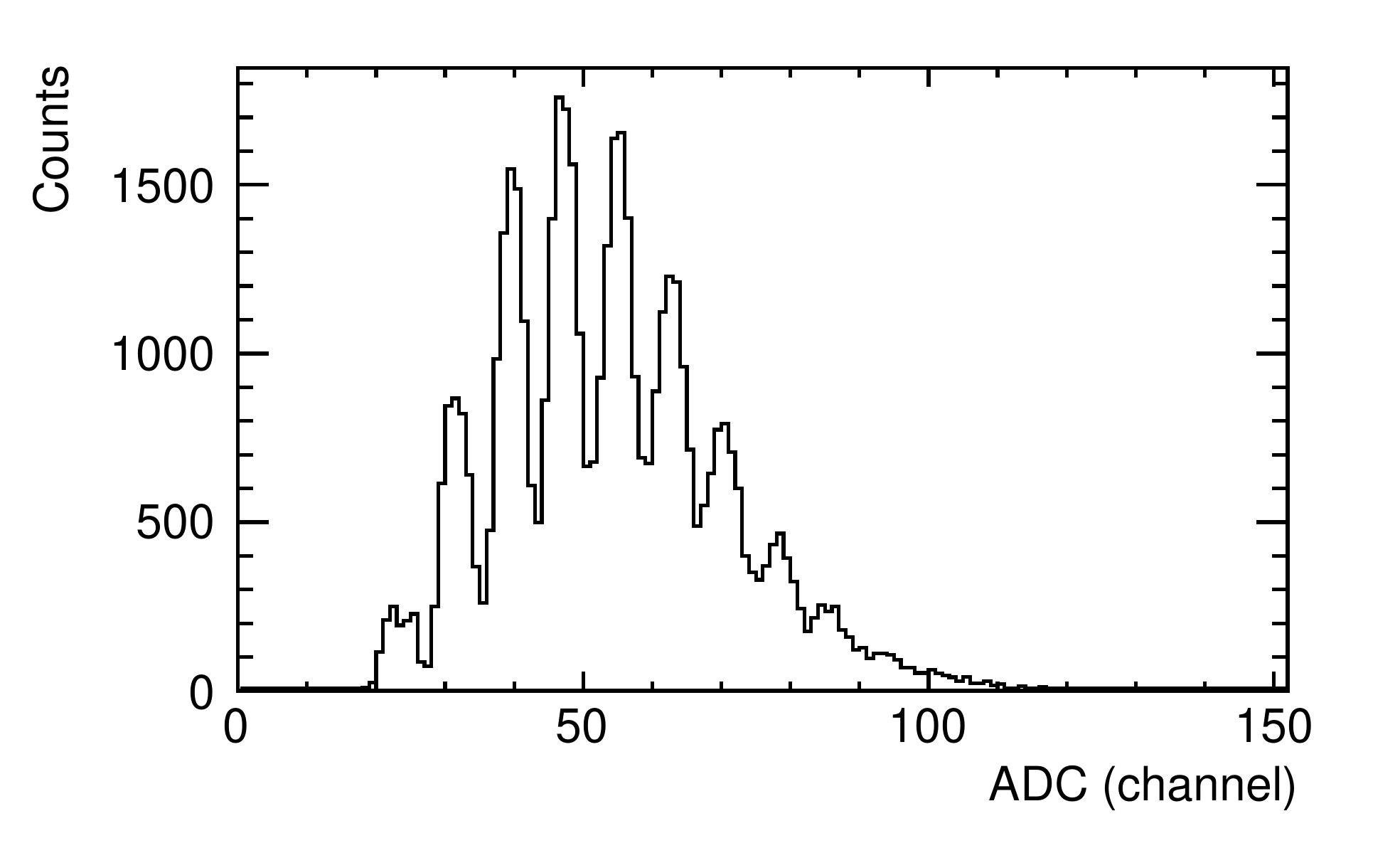}
\caption{\label{fig:photons}ADC spectrum of a SiPM illuminated by LED pulses.}
\end{figure}

Figure \ref{fig:photons} shows an example of the measured ADC spectrum from a SiPM. The peaks in the spectrum correspond to integer numbers of pixels that fire. When the light intensity is low the \emph{pedestal} peak, corresponding to the events in which no pixels have fired, is clearly visible. The center of the pedestal peak, representing the zero signal, and the average distance between the centers of the neighboring peaks, representing the single pixel signal amplitude, are used to express the signal amplitude in terms of the number of pixels that have fired, $n_{pixel}$. 

Each detected photon may cause 1 or more pixels to fire. The ratio of the average number of fired pixels, $\left < n_{pixel} \right >$, to the average number of detected photons, $\nu$, at low light intensity is generally larger than 1 due to the optical and electrical cross-talks, as well as the afterpulsing (see \cite{Dolgo06}). For simplicity we combine all these effects under the common \emph{cross-talk factor} $X = \left < n_{pixel} \right > / \nu$.  

The probability for detecting zero photons is $P_0 = e^{-\nu}$. $P_0$ is measured as the ratio of the number of events in the pedestal peak, $A_0$, to the integral of the whole spectrum, $A$, allowing to extract the average number of detected photons as $\nu = -\ln(A_0 / A)$. On the other hand, the average number of fired pixels is determined from the mean of the measured spectrum. Thus the cross-talk factor is obtained as,

\begin{equation}
\label{eq:x-talk}
X = - \frac{ \left < n_{pixel} \right > }{ \ln(A_0 / A) }
\end{equation}

The cross-talk factor depends on several parameters, including the bias voltage applied to the SiPM and its temperature. The bias voltage was kept constant to within 0.1~V for the duration of the tests. The experimental room was climatized, limiting the temperature variations. During the tests, the value of the cross-talk factor was measured several times, and the results were centered around $X = 1.35$ with relative variations within $\pm 5\%$. 

%% file: Configurations.tex
Four configurations of the scintillator strip with WLS fibers were tested for the light yield. The light yields given here refer to the average number of detected photons per cosmic-ray muon hit. The position of muon hits was distributed in the region of $\pm 30 \unit{cm}$ from the center of the tested strip (see description of the modified setup in Sec.\ \ref{sec:setup}).

\begin{figure*}
  \centering
  \begin{subfigure}{0.42\textwidth}
    \includegraphics[width=\textwidth]{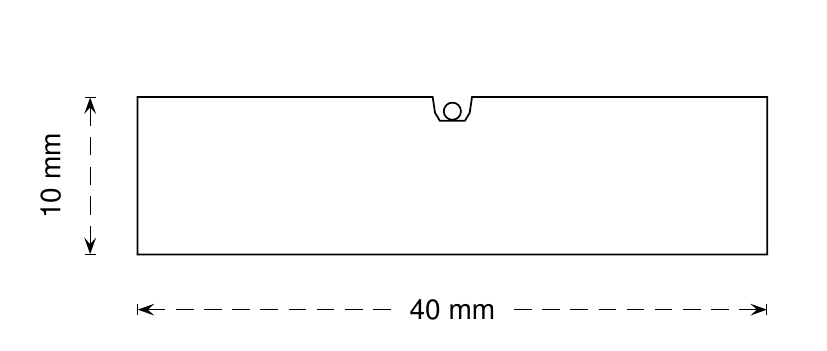}
    \subcaption{\label{fig:confA}}
  \end{subfigure}
  \begin{subfigure}{0.42\textwidth}
    \includegraphics[width=\textwidth]{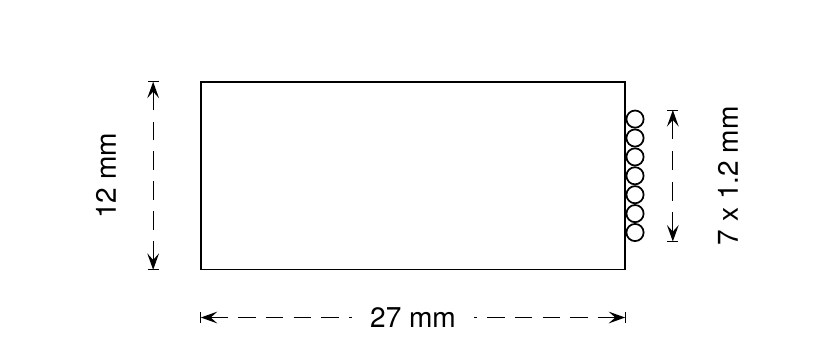}
    \subcaption{\label{fig:confB}}
  \end{subfigure}
  \begin{subfigure}{0.42\textwidth}
    \includegraphics[width=\textwidth]{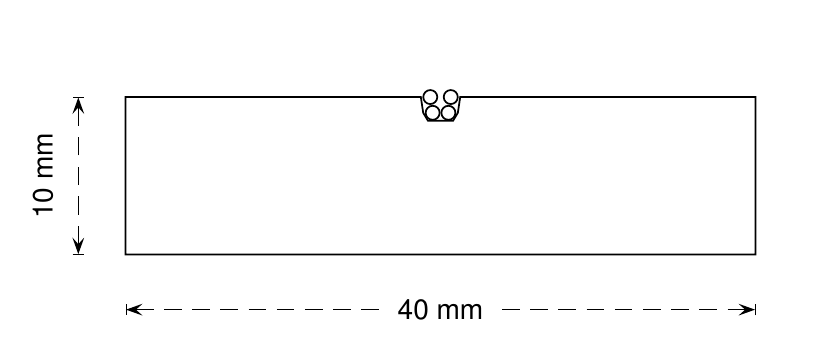}
    \subcaption{\label{fig:confC}}
  \end{subfigure}
  \begin{subfigure}{0.42\textwidth}
    \includegraphics[width=\textwidth]{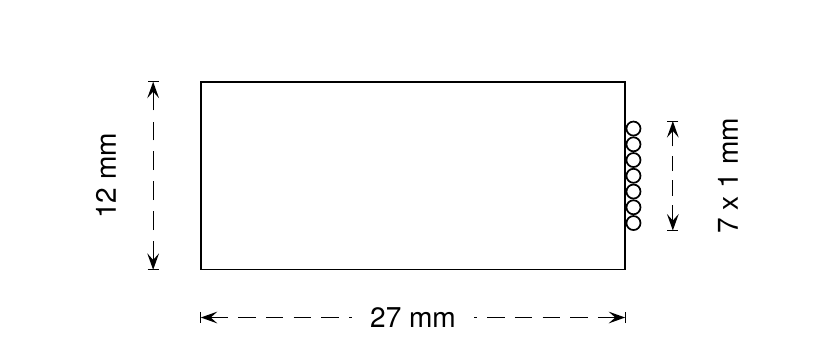}
    \subcaption{\label{fig:confD}}
  \end{subfigure}
  \caption{\label{fig:configurations} Tested configurations of 
                  the scintillator strip with WLS fibers: 
                  \subref{fig:confA} \cfa\ -- 
                         MINOS strip with one Kuraray WLS fiber,
                  \subref{fig:confB} \cfb\ -- 
                         Bicron strip with 7 Kuraray WLS fibers,
                  \subref{fig:confC} \cfc\ -- 
                         MINOS strip with 4 Bicron WLS fibers,
                  \subref{fig:confD} \cfd\ -- 
                         Bicron strip with 7 Bicron WLS fibers.
          }
\end{figure*}

The four configurations, denoted $A$, $B$, $C$ and $D$, are schematically presented in Fig.\ \ref{fig:configurations}. The description of the used strips, fibers and light insulation follows:

\begin{description}
\item{Scintillator strips} \hfill \\
Configurations $A$ and $C$ were built with clear polystyrene scintillator strips with a $40\times10\unit{mm}^2$ profile, co-extruded with a $\text{TiO}_2$ loaded surface layer such as used by the MINOS collaboration \cite{MINOS_sci_08}. Configurations $B$ and $D$ were built with clear \bicron fast scintillator strip with a $27\times12\unit{mm}^2$ profile \cite{BicronStrip}. 
\item{WLS fibers} \hfill \\
Configurations $A$ and $B$ used polystyrene double-clad fibers of 1.2~mm diameter with 175 ppm of Y-11 fluor produced by Kuraray Inc.\ Japan. One such fiber was inserted into the groove and covered with white \tyvek sheet type 1056D \cite{dupont, Abazov05} in configuration $A$, while seven fibers were attached to the narrower side of the strip using reflective adhesive tape in 5 points along the strip in configuration $B$. Configurations $C$ and $D$ used \bicronwls WLS fibers of 1.0~mm diameter. Four fibers were inserted into the groove of the strip in configuration $C$ and covered with the \mbox{\tyvek} sheet, while in configuration $D$ seven fibers were attached to the narrower side of the strip using reflective adhesive tape. Configurations $B$ and $D$ were additionally wrapped with one layer of the \tyvek sheet, and all configurations were finally wrapped in several layers of black \tedlar paper \cite{dupont, Abazov05}. 
\end{description}

In configurations using seven fibers, the ends of the fibers extended 20~cm beyond the end of the strip and were bundled together so that six fibers surrounded one in a tight hexagonal shape. In the configuration using 4 fibers, the fibers were bundled in an approximately square form. In this way it was possible to efficiently collect light from the fibers onto the $3\times3\unit{mm}^2$ area of the SiPMs. No optical glues or greases were used for the connection of the fibers to the scintillator strips or to the SiPMs. Configurations $B$ and $D$ were designed based on the experience from the upgrade of the \dzero muon system  \cite{Evdoki98}.

The measured light yield per muon was 10, 19, 20 and over 30 photoelectrons on each end of the strip in configurations $A$, $B$, $C$ and $D$, respectively. The configurations $C$ and $D$ were selected for detailed position- and time-resolution studies.

%% file: AnalysisNIM.tex
\subsection{Event selection}
\input{Selection_1.tex}

\subsection{Longitudinal position resolution and the speed of signal propagation along the strip}
\label{sec:position1}
\input{Position_1.tex}

\subsection{Time resolution}
\label{sec:time1}
\input{Time_1.tex}

\input{Attenuation.tex}

%% file: Selection_1.tex

The off-line event selection was performed as follows:

\begin{enumerate}

  \item Events in which either of the \sfour, \mbox{\sipmo} or \mbox{\sipmt} signals is below the discriminator threshold, signalled by the end-of-scale value in the respective TDC channel, were rejected. 

  \item Events with energy deposit in the \mbox{\sfour} counter below \mbox{75\%} of the most probable energy deposit were rejected. 

\end{enumerate}

%% file: Position_1.tex
The location of the muon impact was defined by the position $x$ of the \sfour counter along the axis of  the tested strip. The center of the tested strip was assigned the relative position $x=0$, and the $x$ axis was oriented away from \sipmt towards \sipmo. Five points along the strip were measured for the configurations $C$ and $D$. The data for each point were collected over 7 to 12 hours in order to collect the statistics of at least 500 events remaining per point after the selection cuts.

The observable with the best sensitivity to muon position along the strip is the time difference between the two SiPMs. Position is thus measured as,

\begin{equation}
\label{eq:sipmx}
   x_{SiPM} = b_0 + v^* \frac{t_{SiPM_2} - t_{SiPM_1}}{2} = b_0 + v^* \frac{\Delta t}{2}
\end{equation}

where $b_0$ is the offset and $v^*$ is the speed of the signal propagation along the tested strip. 

The time of the rising edge detection by the constant threshold discriminator depends on a signal amplitude. Beside worsening the time resolution, this effect also introduces a position-dependent bias especially at the strip ends due to a difference in SiPM signals caused by different attenuation in the opposite directions.

\begin{figure}
  \centering
  \includegraphics[width=\singleplotwid]{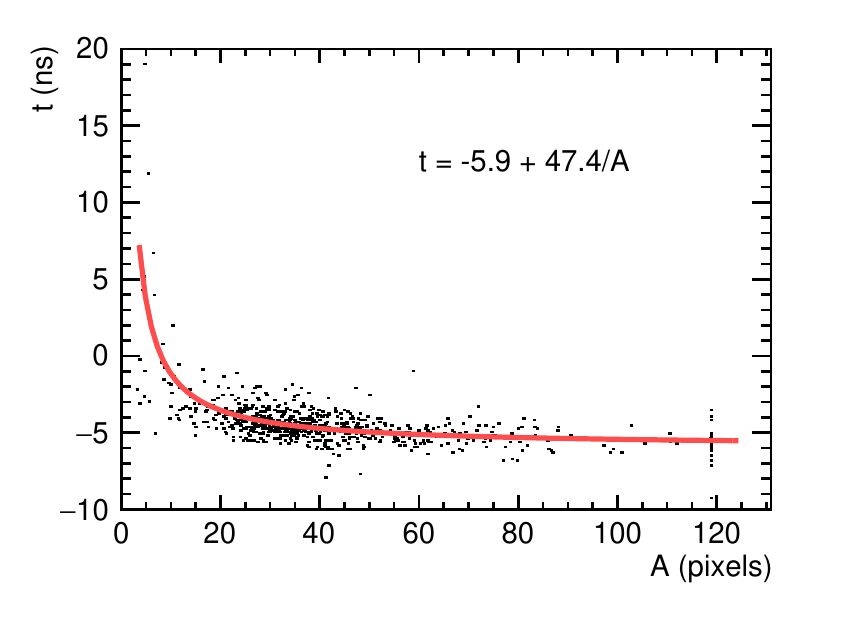}
  \caption{\label{fig:aCorr} (Color online) Scatter plot of time vs.\ amplitude 
                  of \sipmo from the
                  dataset taken at the \sfour position $x=0$ with \cfd, showing 
                  the amplitude-dependent delay of the timing signal from the 
                  discriminator. The function used for the correction 
                  of this effect is also shown (continuous line).}
\end{figure}

The amplitude effect was corrected by subtracting the amplitude-dependent delay of the form $\delta t = a_1/A$, where $A$ is the amplitude of the signal. To obtain the parameter $a_1$, the function $t = a_0 + a_1/A$ was fitted to the scatter plot of the time versus amplitude for both SiPMs separately at $x=0$, i.e.\ when \mbox{\sfour} is at the center of the tested strip (Fig.\ \ref{fig:aCorr}). The parameter $a_0$ is the constant time offset of the SiPM in the high-amplitude limit. \footnote{Correction of the amplitude effect for the \mbox{\sfour} counter, although in principle relevant, was not necessary due to the fast signal and high amplitude in the counter for the selected events.}

\begin{figure}
\centering
\includegraphics[width=\singleplotwid]{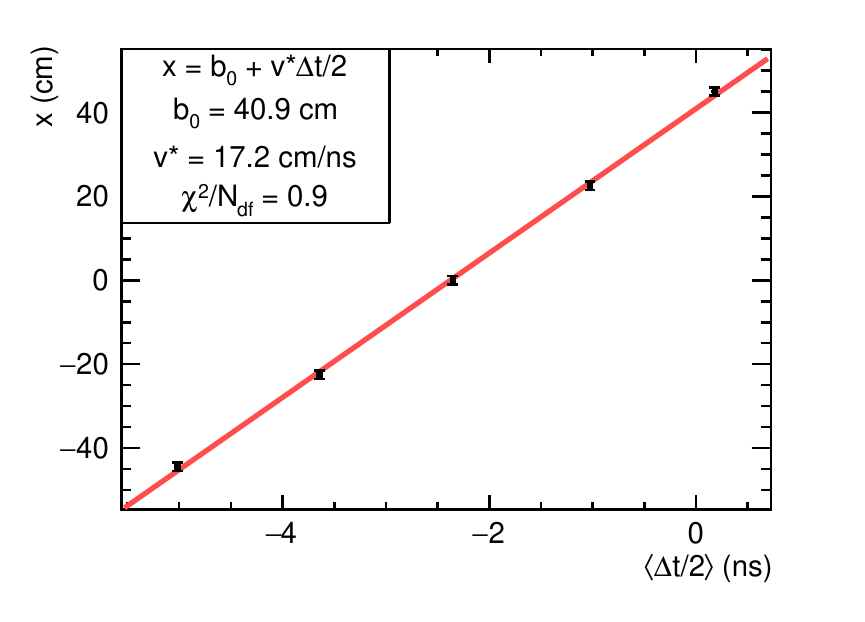}
\caption{\label{fig:speed1} (Color online) Plot of $x$ vs. 
                   $\frac{\Delta t}{2}$ for the \cfd. Linear fit is also shown.}
\end{figure}

The speed of the signal propagation along the strip is determined from the linear fit to the plot of $x$ vs.\ $\left< \Delta t/2 \right>$ in the five measured points, as shown in Fig.\ \ref{fig:speed1}. The uncertainty on $\Delta t/2$ was estimated from the scatter of the data, while the uncertainty on $x$ was set to 1~cm, corresponding to the estimated precision of the position of the \sfour counter. 

\begin{figure}
\centering
\includegraphics[width=\singleplotwid]{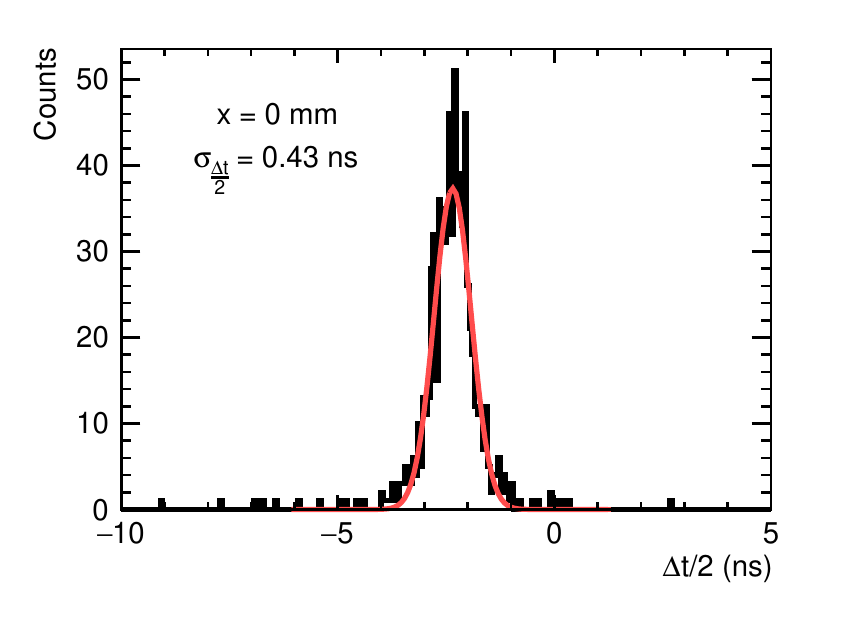}
\caption{\label{fig:x0} (Color online) Distribution of $\Delta t/2$ after the amplitude correction for the \cfd and $x=0$. Gaussian fit is also shown.}
\end{figure}

The distribution of the variable $\Delta t/2$ after the amplitude correction (Fig.\ \ref{fig:aCorr}) is shown in Fig.\ \ref{fig:x0} for the \cfd and $x=0$. The standard deviation of the distribution, determined from a Gaussian fit, is the crucial parameter for the counter position resolution. The average value of the standard deviation of $\Delta t/2$ from all 5 measured points in \cfd was $\sigma_{\Delta t/2} = \besttime$.

The position resolution along the strip can be estimated as $\sigma_x = \sigma_{\Delta t/2} v^* = \bestpos$ for the \cfd. Beside the position resolution of the tested strip, this estimate contains contributions from the uncertainty of the muon impact position along the tested strip due to the width of the \sfour counter and the uncertainty due to the angular distribution of the muon tracks across the distance between \sfour and the tested strip. This estimate of the coordinate resolution along the tested counter can thus be regarded as conservative.

%% file: Time_1.tex
\begin{figure}
\centering
  \includegraphics[width=\singleplotwid]{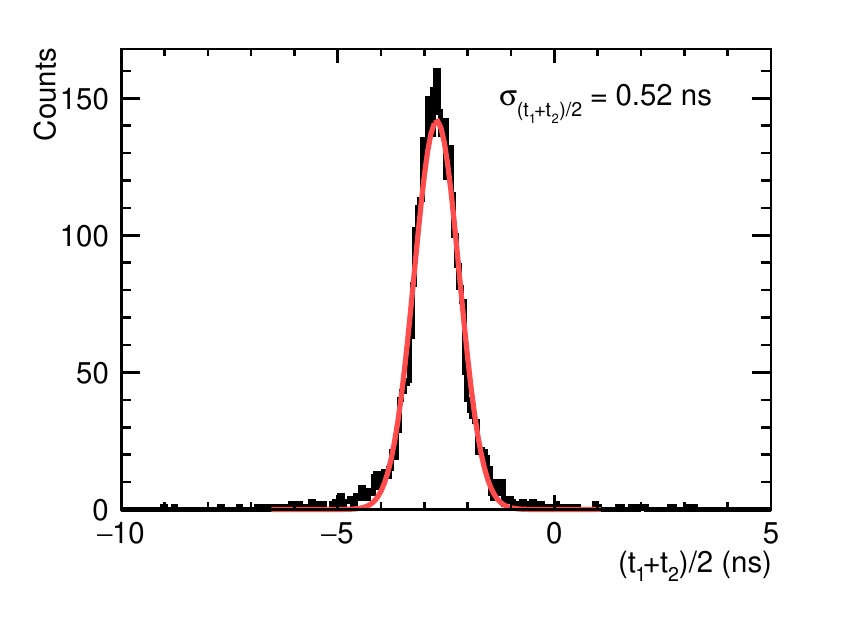}
  \caption{\label{fig:time1} (Color online) Distribution of the average times for 
              the strip \cfd w.r.t.\ \sfour. 
              Gaussian fit is also shown.}
\end{figure}

The most direct way to measure strip time resolution is to analyze the 
distribution of the average time of 
the two SiPMs at the ends of the tested strip $(t_1+t_2)/2$. 
As the average time of the tested strip does not depend on 
the position of \sfour along the tested strip within measurement uncertainty, 
the data from all 5 studied \sfour positions are added up.
The distribution of average times is shown in Fig.\
\ref{fig:time1} for the \cfd. The standard deviation of the fitted 
Gaussian curve is 0.52~ns.
The dominant contribution to the width of the distribution is the time 
resolution of the tested strip, but other contributions are also present, 
such as the time resolution of \sfour. Also, the width of the tested strip 
introduces a spread in the muon positions w.r.t.\ the PMT of \sfour.

Another estimate of the time resolution of the tested strip can be 
inferred from $\sigma_{\Delta t/2}$. Since amplitude effect on $t_1$ and $t_2$ is corrected as shown in Fig.\ \ref{fig:aCorr}, the statistical fluctuations 
of $t_1$ and $t_2$ are independent of each other and the variance of 
$\Delta t/2 = (t_1-t_2)/2$ is the same as that of $(t_1+t_2)/2$.
The average value of 
$\sigma_{\Delta t/2}$ for the 5 measured points in \cfd is 0.45~ns. 
The time resolution of \sfour does not contribute to $\sigma_{\Delta t/2}$. 
Nevertheless, contributions from uncertainties other than the time resolution 
of the tested strip are still present, such as the uncertainty of the muon-
position due to the width of \sfour counter.

The time resolution of both SiPM channels individually was determined as the average Gaussian width of the distribution of $t_{\sipm} - t_{\sfour}$ in all five measured points. In \cfd the result is $\sigma_{t,\sipmo} = 0.68\unit{ns}$ and $\sigma_{t,\sipmt} = 0.63\unit{ns}$.

%% file: Attenuation.tex
\subsection{Attenuation of light along the strip}
\label{sec:attenuation}

Attenuation of the light signal along the tested strip was measured in the modified setup described in \mbox{Sec.\ \ref{sec:setup}}. 

\begin{figure}
\centering
   \includegraphics[width=\singleplotwid]{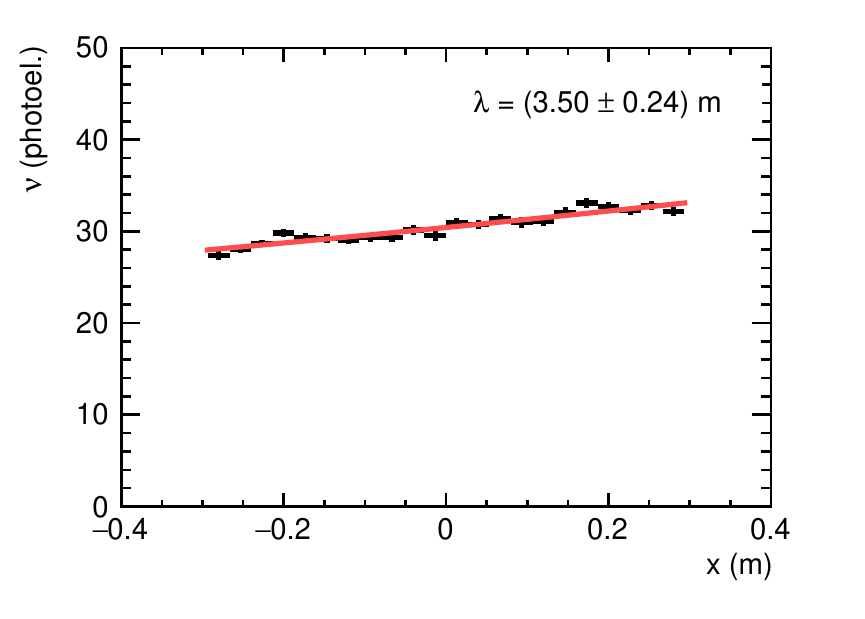}
   \caption{\label{fig:attenuation} (Color online) Plot of the mean value of the 
               \sipmo photon count vs.\ muon position in \cfd. Fit of the exponential 
               function and the attenuation length are also shown.}
\end{figure}

Figure \ref{fig:attenuation} shows a plot of the mean value of \sipmo photon count in \cfd vs.\ muon position. Exponential fit to the data indicates an attenuation length for the light signal in this strip of $\lambda = 3.5 \pm 0.2 \unit{m}$. This value is consistent with the statement of the manufacturer of the WLS fibers that $\lambda \ge 3.5 \unit{m}$ \cite{BicronWLS}. 

The observed attenuation length of the tested strips is smaller than the expected size of the muon system. Increase in both the attenuation length and the light yield will be beneficial for the muon system design.

%% file: Results.tex
\section{Results}
\label{sec:results}

\begin{table}
\caption{\label{tab:sipm} Measured performance of the individual SiPM readout 
   channels for the configurations $C$ and $D$.}
\centering
	\begin{tabular}{ c | c | c | c }
    \hline
	Configuration  & SiPM  & Light yield / muon & $\sigma_{t,\sipm}$ \\
		           &   \#  &  (photoelectrons)  &  (ns)  \\
	\hline
	\multirow{2}{*}{$C$} & 1 & 21 & 1.2 \\
                             & 2 & 20 & 1.1  \\
    \hline
	\multirow{2}{*}{$D$} & 1 & 31 & 0.68 \\
	                     & 2 & 36 & 0.63 \\
    \hline
	\end{tabular}
\end{table}

\begin{table}
\caption{\label{tab:results} Measured properties of the strip configurations 
        $C$ and $D$.}
\centering
	\begin{tabular}{ c | c | c | c | c  }
    \hline
	Configuration  & $\sigma_x$ & $\sigma_{t,strip}$ &  $v^*$  & $\lambda$\\
		           &   (cm)     &         (ns)       & (cm/ns) &   (m)    \\
	\hline
	$C$            &   16.0     &         0.88       &    18.1 &   2.1    \\
	$D$            &   7.7      &         0.45       &    17.2 &   3.5    \\
    \hline
	\end{tabular}
\end{table}

Table \ref{tab:sipm} summarizes the measured performance of the individual SiPM readout channels for the configurations $C$ and $D$. The average number of photoelectrons per muon is sufficiently high to ensure a 100~\% detection efficiency for muons traversing the entire thickness of the strip. Time resolution per readout channel of the order of 1~ns or better was achieved with signals of only 20 to 30 photoelectrons on average.

Table \ref{tab:results} summarizes the results of the strip studies including position and time resolutons, signal propagation speed and the attenuation length. The difference in the resolutions for the two configurations is larger than the ratio expected if the resolution follows the $1/\sqrt{\nu}$ law, where $\nu$ is the number of photons. This indicates that other factors besides the statistical effect of the photon yield influence the time and the position resolutions. These additional effects include the properties of the scintillator and the WLS fiber materials, especially the light emission time. 

The best measured position resolution in the studies described here is \bestpos, achieved with the \cfd.

%% file: Conclusions.tex
\section{Conclusions}
\label{sec:conclusions}

Prototype scintilator+WLS strip configurations with SiPM readout for a muon system for the future colliders were tested for light yield, position resolution and time resolution. Depending on the configuration, a light yield in single SiPM of up to 36 photoelectrons per muon has been achieved. Strip time resolution of \besttime and position resolution of \bestpos were achieved. Tests with higher statistics more precise timing and muon position reference, such as in the test-beam, will yield results with better accuracy.

A muon system for future colliders based on scintillator strips with WLS fibers and SiPM readout would possess excellent muon-detection efficiency, a time resolution suitable even for challenging beam time structure such as that of the CLIC design, and a position resolution sufficient for track matching with the central detector for the broad range of physics processes. The attenuation length of the WLS fibers limits the acceptable strip length. In the endcap region of a CLIC collider, the occupancy from muon beam halo background also presents a constraint on the strip length \mbox{\cite{Kraaij11}}.
The reliability of operation, the economy of production and the relatively low required number of readout channels make this technology a very attractive option.

%% file: Acknowledgments.tex
\section{Acknowledgments}
\label{sec:acknoledgments}

The authors acknowledge the support received from the Ministry of Education and Science and the National Research Center ``Kurchatov Institute'' (Russian Federation), from the Ministry of Education, Science and Technological Development  (Republic of Serbia) within the project OI171012 and from the Department of Energy (United States of America).